# X-ray diffraction study of epitaxial CuO nanostructures obtained through post-deposition annealing of Cu on SrTiO$_3$(001)

Maurizio De Santis[a*], Veronique Langlais[b], Lucio Martinelli[a], Tom Mocellin[a], Sébastien Pairis[a], Xavier Torrelles[c]

[a] Université Grenoble Alpes, CNRS, Grenoble INP, Institut Néel, BP 166, 38042 Grenoble, France

[b] CNRS, CEMES (Centre d'Elaboration des Matériaux et d'Etudes Structurales), BP 94347, 29 rue Jeanne Marvig, F-31055 Toulouse, France

[c] Institut de Ciència de Materials de Barcelona, ICMAB-CSIC, Bellaterra, 08193 Barcelona, Spain

* Corresponding author. E-mail address: Maurizio.de-santis@neel.cnrs.fr

**ABSTRACT**

Orientation, structure and morphology in the early growth stages of CuO films on strontium titanate were studied by synchrotron radiation X-ray diffraction. Nanostructured CuO films were obtained by *ex situ* heat treatment at 970 K in oxygen flow at ambient pressure after an initial deposition of Cu on SrTiO$_3$(001) at 10$^{-4}$ Pa oxygen pressure using the molecular beam epitaxy technique. These films at low coverages grow forming islands a few tens of nanometers wide, with the [010] direction perpendicular to the substrate surface. Two types of epitaxies are observed, CuO[001]//SrTiO$_3$[100] and CuO[100]//SrTiO$_3$[100], the former being the preferred one. The combination of these epitaxies with the P4mm surface symmetry of the substrate generates 16 different orientations of nanostructures. The lattice constant values obtained from X-ray diffraction are very close to those of bulk tenorite, with the exception of the monoclinic angle $\beta$, which is reduced from 99.54° to 96.76(8)° by the epitaxy constraints. This angle is a key parameter in determining the magnetic properties of CuO.

1. Introduction

The tenorite CuO phase is unique among the monoxides of the 3$d$ transition elements having a monoclinic unit cell and square planar coordination of copper by oxygen, rather than the cubic rock salt structure and octahedral coordination. At room temperature, it is paramagnetic and presents two ordered magnetic phases at low temperature. Below 213 K, it is antiferromagnetic and its spin structure is collinear with a commensurate propagation vector (1/2 0 -1/2) and spins aligned along the **b** direction (AF1 phase). Above this temperature, it shows an incommensurate antiferromagnetic structure, which persists up to 230 K (AF2 phase) [1]. A renewed interest in the CuO properties has arisen since it was discovered that the latter phase is a type-II multiferroic, wherein ferroelectricity is induced by non-collinear spiral magnetic order breaking the inversion symmetry. In CuO this gives rise to a polarization of more than 100 µC/m² along the **b** axis [2]. The strong magnetoelectric coupling in type-II multiferroics opens the way to the design of multifunctional devices, where charging is driven by applying an external magnetic field or, vice versa, spin state is controlled through applied voltages. Unfortunately, most of the known induced-multiferroics exhibit low ordering temperatures (below 40 K), which is a serious drawback for their application in devices. The outstanding feature of CuO is its surprisingly high ferroelectric Curie temperature ($T_C$) compared to the others multiferroics in this class. This make of tenorite a potential candidate for the insertion in technological devices.

Tenorite has $a$=468.37(5) pm, $b$=342.26(5) pm, $c$=512.88(6) pm, and $\beta$=99.54(1)° as lattice constants [3]. Its structure forms two types of zigzag Cu–O chains running along the [10-1] and [101] directions, respectively. The Cu-O-Cu bond angle $\theta$ is directly related to the superexchange interaction $J$ [4]. The mechanism that stabilizes multiferroicity in CuO is still not fully understood, and remains the subject of significant debate. Theoretical calculations highlighted either the role of superexchange interaction [5], or the strength of interactions between Cu-O chains [6], or again a crucial role of the Dzyaloshinskii-Moriya interaction [7]. In any case, all modifications of Cu-O-Cu bonding angles and distances can modify $T_C$ and possibly increase it up to room temperature. Indeed, theoretical calculations predict that application of high hydrostatic pressure onto CuO results in a huge stabilization of the multiferroic phase, extending its stability range well above room temperature [8].

Epitaxial growth also induces stress in CuO films. The growth of high quality CuO films in vacuum conditions compatible with molecular beam epitaxy (MBE) is a challenge, since at high temperature and low molecular oxygen pressure CuO decomposes resulting in $Cu_2O$ [9]. Indeed, an activated form of oxygen, such as atomic oxygen or ozone, is required. Epitaxial CuO was obtained in MBE regime by depositing Cu in atomic oxygen generated by a RF plasma source onto MgO(100) and Y-stabilized $ZrO_2$(100), resulting in CuO(111) and CuO(010) films, respectively [10,11]. More recently, CuO tenorite thin films were grown using Pulsed Laser Deposition (PLD) technique with 30 Pa partial oxygen pressure [12]. The intercalation of a MgO buffer layer resulted in CuO(111) films on $SrTiO_3$(001), and in CuO(010) ones on $SrTiO_3$(110). PLD was also employed by another group to grow a new tetragonal CuO phase, in coherent epitaxy on SrTiO$_3$(001), up to a thickness of about 3 nm [13]. This new phase has unusual properties, due to the Cu-O-Cu bonding angle of 180°.

In this study, CuO tenorite films were epitaxially grown in two steps. First, thin epitaxial Cu films were deposited on SrTiO$_3$(001) in molecular oxygen at a partial pressure of 10$^{-4}$ Pa by MBE. Then, *ex situ* annealing in O$_2$ was performed to obtain CuO tenorite phase. The structure of films was investigated, before and after post annealing, by x-ray diffraction (XRD) using synchrotron radiation. Post-deposition annealing in air or oxygen is a standard method to obtain oxide films, which are in most cases polycrystalline (see e.g. [14], [15]). It seldom results in epitaxial oxides, like e.g. for epitaxial growth of VO$_2$ through post-deposition annealing [16].

## 2. Experimental details

### 2.1. Experimental set-up

The present experiment was performed at the European Synchrotron Radiation Facility (ESRF, Grenoble, France) using the dedicated In Situ Nanostructures and Surfaces (INS2) apparatus of the BM32 beamline. The X-ray source is a bending magnet and the monochromator is a Si(111) double crystal, with the second crystal bent to give sagittal focusing on the sample. The vertical focusing is provided by two mirrors, resulting in a spot size of about 0.2×0.3 mm$^2$ at the sample position. The experimental station consists of an ultrahigh vacuum (UHV) chamber with a base pressure in the low 10$^{-8}$ Pa range and equipped for sample preparation and analysis. This includes sample annealing, a leak valve for O$_2$ dosing, evaporators, and Auger Electron spectroscopy (AES) analysis technique. This chamber has two beryllium windows, a flat one and a large cylindrical one, for the incoming and outgoing X-ray beam. It is mounted on a z-axis diffractometer [17]. A hexapod installed on the sample azimuthal axis allows the alignment of the surface normal parallel to it and the sample placement at the center of the diffractometer. The measurements were performed at a photon energy of 14 keV to avoid the fluorescence contribution from Sr *K*-edge absorption. The surface morphology was investigated at Néel Institute by Scanning Tunneling Microscopy (STM) and Scanning Electron Microscopy (SEM). STM was performed using a UHV set-up composed by a preparation chamber with a suitable oven coupled to an analysis chamber equipped with a commercial STM (Omicron VT STM/AFM). SEM was performed using a Zeiss Ultra+ instrument with a charge compensation device allowing for measurements on insulating samples.

### 2.2. Sample preparation

The substrate was a polished (001) oriented $SrTiO_3$ single crystal (Crystec GmbH), 0.5 mm thick, with a squared surface of 10×10 mm$^2$ and a miscut of less than 0.1° (STO in the following). It was first sonicated in deionized (DI) water at about 340 K for 20 min, then put in an alumina boat inside a high temperature tube furnace and annealed for 4 hours at 1273 K in an $O_2$ flux at atmospheric pressure, and finally rinsed again in DI water. This preparation procedure is close to that one reported in ref. [18], which resulted in atomically flat and $TiO_2$-terminated surfaces. Afterwards, substrates were introduced in UHV and annealed for 20 min at about 1120 K in $10^{-4}$ Pa $O_2$, to remove surface contaminants. AES shows some residual carbon contamination which was evaluated from the ratio of carbon versus titanium peaks at about 1 C atom per nm$^2$.

Metallic copper was evaporated on the substrate from a water-cooled Knudsen cell, with the chamber backfilled with molecular oxygen at a partial pressure of $10^{-4}$ Pa. Previously, the Cu flux was calibrated using a quartz crystal microbalance. The deposition rate was about 44 pm/min. In the following, a sample with a nominal coverage of 2.2 nm will be discussed in detail. During Cu deposition, the substrate temperature was kept at about 910 K. These conditions are not suitable to form CuO, which would decompose following the Cu-O phase diagram [9] and were set to grown an oxide precursor. In the following, we'll refer to this sample as as-deposited. After studying its structure and the epitaxy relationship, the sample was removed from UHV and annealed in the tube furnace for 4 hours at 970 K in $O_2$ at atmospheric pressure, which are conditions within the bulk CuO stability range [19]. Then the sample was inserted back into the INS2 set-up at ESRF and measured again. At this stage, it will be referred to as post-growth annealed sample.

### 2.3. Data acquisition

XRD measurements were performed at room temperature. The data were collected using a 2D detector (MAXIPIX, ESRF), with the beam impinging on the sample at a grazing angle of 0.2°, close to the substrate critical angle for total reflection at 14 keV (except, of course, for the specular reflectivity measurements). The sample was fixed with two tantalum clips welded to the sample holder that, at grazing incidence, shadowed the incoming beam at well-defined azimuths. The data shown in the following are integrated in a region of interest of 4×4 mm² of the detector, corresponding to a resolution of about 0.4 nm$^{-1}$ in grazing incidence and 0.3 nm$^{-1}$ in the specular reflectivity. The momentum transfer is measured in the STO crystallographic axis frame. The relationship between momentum transfer and diffractometer angles is given in ref. [17].

## 3. Results & discussion

### 3.1. Morphology

The surface morphology of control STO substrates that had suffered the treatment described in section 2.2 was studied by STM. The annealing at 1120 K in $10^{-4}$ Pa $O_2$ generates enough oxygen vacancies to evacuate the tunneling charge. Figure 1 reproduces a $400 \times 400$ nm$^2$ STM image collected with a bias sample voltage of 3.2 V, large enough to exceed the gap and inject electrons in the STO conduction band. This picture shows large flat terraces with a few steps, like e.g. the three ones observed along the traced line. All measured steps are 400 pm high, close to the STO lattice constant. This indicates that only one kind of bulk (001) atomic plane is exposed at the surface, in agreement with ref. [18].

Post-growth annealing in oxygen fills STO oxygen vacancies rendering the substrate insulating. STM is then no more feasible, and the CuO film morphology was therefore measured by SEM. The SEM image of Fig. 2 shows a Wolmer-Weber growth mode with the formation of CuO islands of size ranging from about 10 to more than 20 nm.

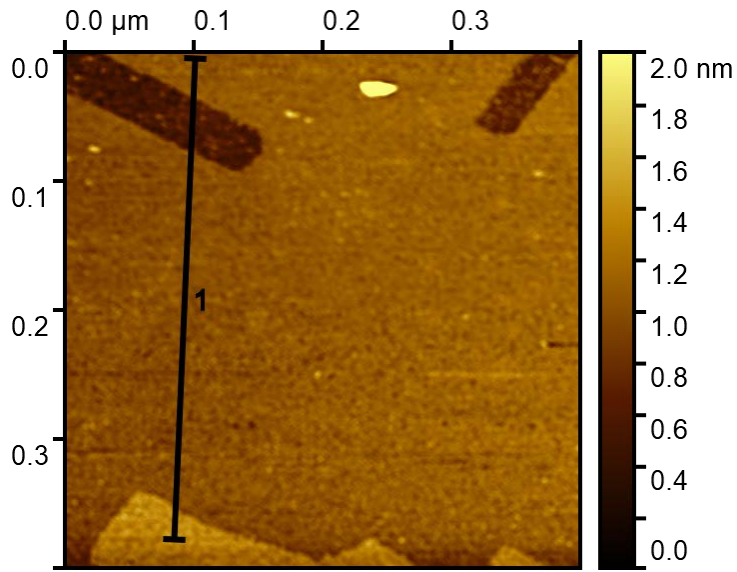

Figure 1. 400×400 nm² STM image of a SrTiO$_3$(001) crystal surface after annealing at 1120 K in 10$^{-4}$ Pa O$_2$ ($V_s$=3.2 V, $I_t$=0.1 nA).

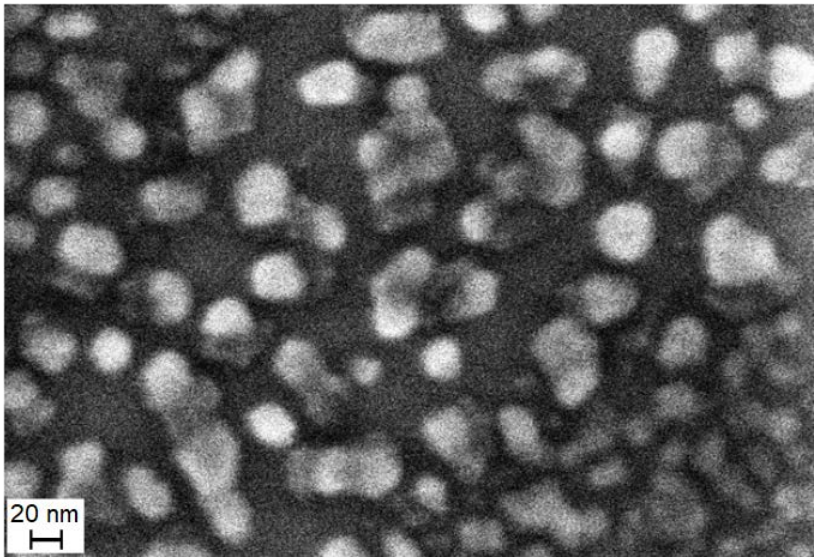

Figure 2. SEM measured ex situ on the post-growth annealed sample (E=2.0 kV).

### 3.2. As-deposited

Figure 3 shows the specular reflectivity in the region close to the STO(002) Bragg peak collected on the as-deposited sample obtained by deposition 2.2 nm of Cu (filled black circles). The measurements show the presence of a strong Cu(002) peak due to a copper film with [001] orientation. A much weaker peak is observed at about 30 nm$^{-1}$, which is attributed to a small amount of oxygen-deficient $Cu_2O$, also with [001] orientation. The film has a thickness of 6.0(1) nm, evaluated by applying the Scherrer equation to the Cu peak width [20]. This is about three times the nominal coverage and is interpreted as a Volmer-Weber growth mode with the formation of islands 6 nm thick separated by uncovered regions. This agrees with Auger, showing Ti peaks that would not be observed on a uniform film of such a thickness.

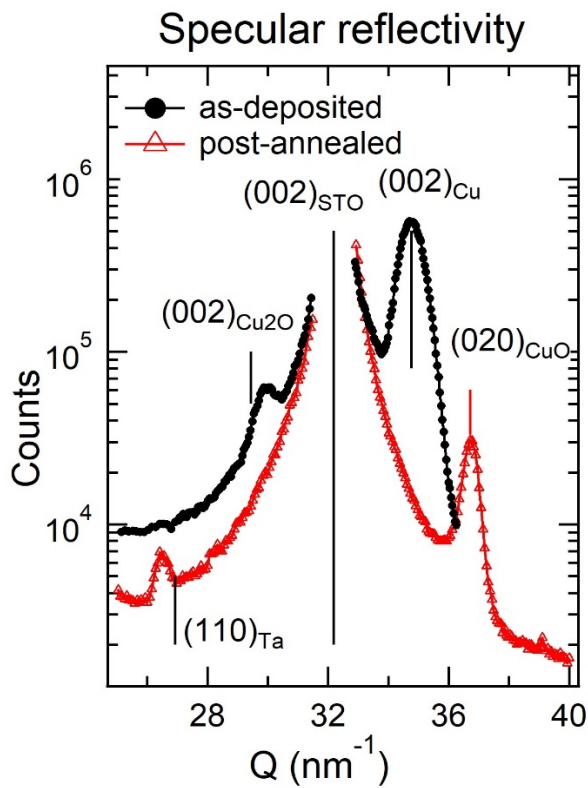

Figure 3. Specular reflectivity versus momentum transfer measured on the as-deposited sample (filled black circles), and on the post-growth annealed sample (empty red triangles). Vertical

*lines represent the bulk values for the indicated STO, Cu, Cu$_2$O and CuO reflections. The clips holding the sample give a small Ta(110) peak at about 27 nm$^{-1}$ (lightly shifted because the clip is not at the center of the diffractometer).*

Figure 4 shows an in-plane scan of the momentum transfer along the x axis (parallel to the STO (100) reciprocal space direction) at $Q_y$=0, $Q_z$=0.5 nm$^{-1}$. The film grows in epitaxy with Cu[100]//STO[100], as evidenced by the presence of the Cu(200) reflection. The mosaicity was measured on a film grown in the same way but with a thickness of 0.55 nm. A rocking scan of the Cu(200) reflection around the surface normal gave in that case about 2°. A small bump close to 30 nm$^{-1}$ is the signature of a minor Cu$_2$O phase. The full-width at half-maximum of the Cu(200) peak is $\Delta Q_x$=0.62 nm$^{-1}$ corresponding to a coherent Cu domain size of 9.5 nm at least, parallel to the surface. This is underestimated due to the convolution of the diffracted beam with the experimental resolution.

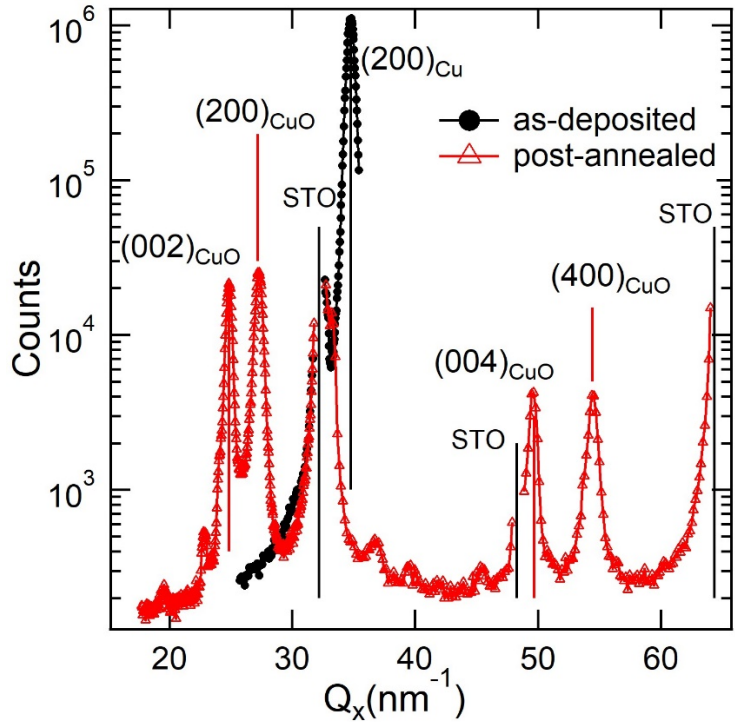

*Figure 4. Scan of the momentum transfer parallel to the STO (100) reciprocal space direction at $Q_y=0$, $Q_z=0.5$ nm$^{-1}$. Filled black circles: as-deposited sample; empty red triangles: post-growth annealed sample. Due to the presence of several domains, both the (002) and (200) reflections of CuO are observed on the post-growth annealed sample (see discussion in section 3.3).*

To crosscheck the film structure, we have finally scanned the momentum transfer perpendicular to the surface, keeping the in-plane component at the Cu(200) value ($Q_x$ =34.83 nm$^{-1}$, see Fig. 5).

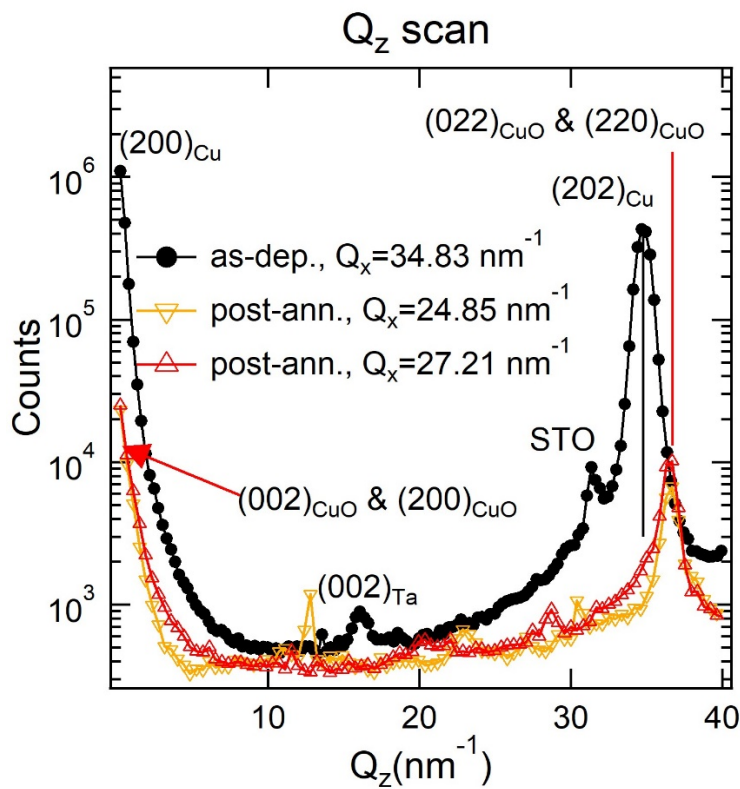

Figure 5. Scan of the momentum transfer perpendicular to the surface, at $Q_y=0$. Filled black circles: as-deposited films; empty red and orange triangles: post-growth annealed sample, measured at two different in plane $Q_x$ values. As in previous figures, vertical lines represent the bulk values for the indicated Cu and CuO reflections. On the as-deposited film, diffuse scattering from the nearby STO(202) reflection is observed at $Q_z=31.3$ nm$^{-1}$.

The scan crosses the Cu(202) reflection at the expected $Q_z$, confirming the film structure. The values of momentum transfer obtained fitting peaks of Figures 3, 4 and 5 are listed in table 1 together with the corresponding bulk values. The as-deposited ones agree very well with the Cu bulk values.

*Table 1*. *Best fit positions of the observed diffraction peaks (using a Gaussian function) and corresponding bulk values*

|  | Film | | | | Bulk* | |
|---|---|---|---|---|---|---|
|  | As-deposited | | Post-growth ann. | | | |
|  | $Q_x$ (nm$^{-1}$) | $Q_z$ (nm$^{-1}$) | $Q_x$ (nm$^{-1}$) | $Q_z$ (nm$^{-1}$) | $Q_x$ (nm$^{-1}$) | $Q_z$ (nm$^{-1}$) |
| Cu(002) |  | 34.756(2) |  |  |  | 34.762 |
| Cu(200) | 34.747(2) |  |  |  | 34.762 |  |
| Cu(202) | - | 34.9(1) |  |  | 34.762 | 34.762 |
| CuO(020) |  |  |  | 36.728(4) |  | 36.716 |
| CuO(002) |  |  | 24.837(3) |  | 24.846 |  |
| CuO(200) |  |  | 27.262(3) |  | 27.206 |  |
| CuO(004) |  |  | 49.59(1) |  | 49.692 |  |
| CuO(400) |  |  | 54.46(1) |  | 54.412 |  |
| CuO(220) |  |  | - | 36.6(1) | 27.206 | 36.716 |
| CuO(022) |  |  |  | 36.7(1) | 24.846 | 36.716 |

* S. Åsbrink, L.-J. Norrby, Acta Cryst. B26 (1970) 8

### 3.3. Post-growth annealed

The film discussed in the previous section was investigated once again after *ex situ* annealing, resulting in a less trivial epitaxial relationship. A peak is now observed along the specular reflectivity at about 36.7 nm$^{-1}$ (red triangles in Fig. 3), which corresponds to the CuO(020) reflection (Table 1). The film thickness is increased to about 15 nm as found from the peak width according to Sherrer equation. This suggests that coalescence happens during high-temperature oxidation, with a rearrangement of deposited Cu atoms to form larger islands. The formation of an epitaxial CuO film is confirmed by in-plane scans. Fig. 4 shows that, along the STO (100) reciprocal space direction, both the CuO(002) and the CuO(200) reflections are observed, together with the respective next order ones, CuO(004) and CuO(400) (see Table 1). This in-plane scan involves several domains. The average domain size parallel to the surface obtained from the peak's width is about 15 nm, in agreement with SEM. The domain structure was investigated by mapping a large portion of the reciprocal space parallel to the surface, at $Q_z=0.5$ nm$^{-1}$, with Q between 13.3 and 30 nm$^{-1}$ and over 200° of azimuthal angle. This reciprocal space map is reported in Fig. 6. It reflects the *P4mm* substrate symmetry, apart from some glitches close to (15 -25 0.5) nm$^{-1}$ that originate from the sample holder and a shadowing effect of the Ta clips holding the sample at about (3 25 0.5) nm$^{-1}$. The substrate contribution is observed at (16.09 ±16.09 0.5) nm$^{-1}$, which corresponds to the (1 1) and (1 -1) STO crystal truncation rods.

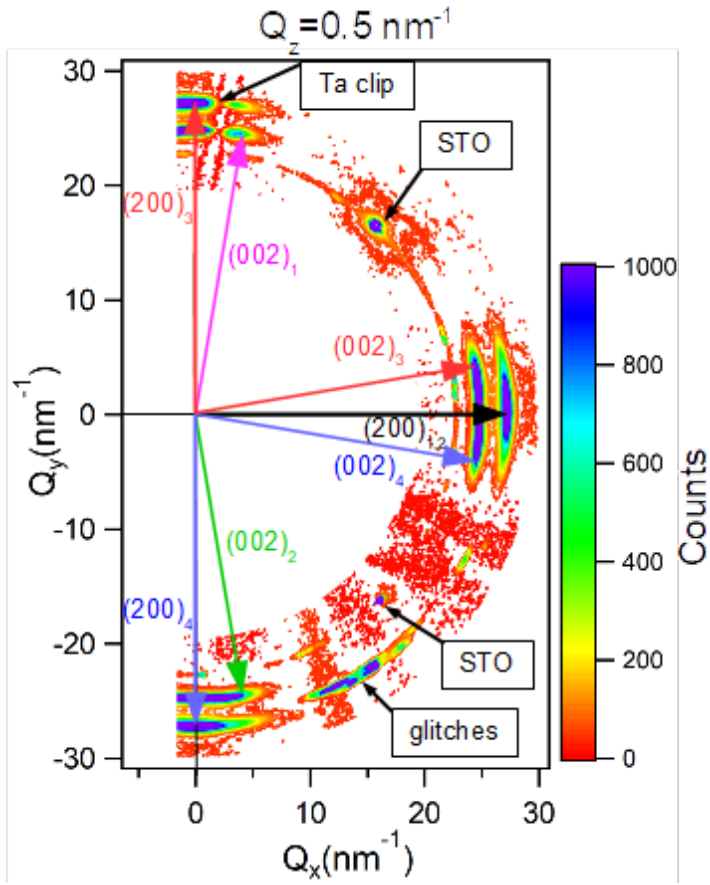

Figure 6. In-plane reciprocal space map collected on post-growth annealed sample at $Q_z=0.5$ nm$^{-1}$ with $13.3 < Q < 30$ nm$^{-1}$ and over 200° of azimuthal angle.

All other features can be assigned to the CuO(002) and CuO(200) reflections. This map shows that, at each (200) reflection can be associated a (002) reflection in the surface plane belonging to the same domain. In other words, all domains have the [010] direction perpendicular to the surface. This is coherent with the reflectivity measurement and is confirmed by the out-of-plane measurements of Fig.5. The CuO(200) reflection is aligned with the STO (100) reciprocal space direction. This corresponds to an epitaxial relationship with CuO[001]//STO[010]. The lattice matching is $(3\times c_{CuO}-4\times a_{STO})/(4\times a_{STO})=-1.5\%$. Then the CuO [100] direction should be 9.54° out of the substrate crystallographic axes, according to the CuO bulk structure, resulting in 8 domains in the *P4mm* symmetry. Four of them are represented in figure 6 (the others are obtained by mirror symmetry with respect to the STO [010] axis). At first sight, this explains the large angular distribution of the CuO(002) reflection. However, a more careful analysis shows a significant discrepancy of the CuO epitaxial film structure with respect to the bulk one. Fig. 7 shows two azimuthal scans performed at Q=24.84 and Q=27.26 nm$^{-1}$, corresponding respectively to the CuO(002) and CuO(200) reflections, and centered on the STO [100] axis.

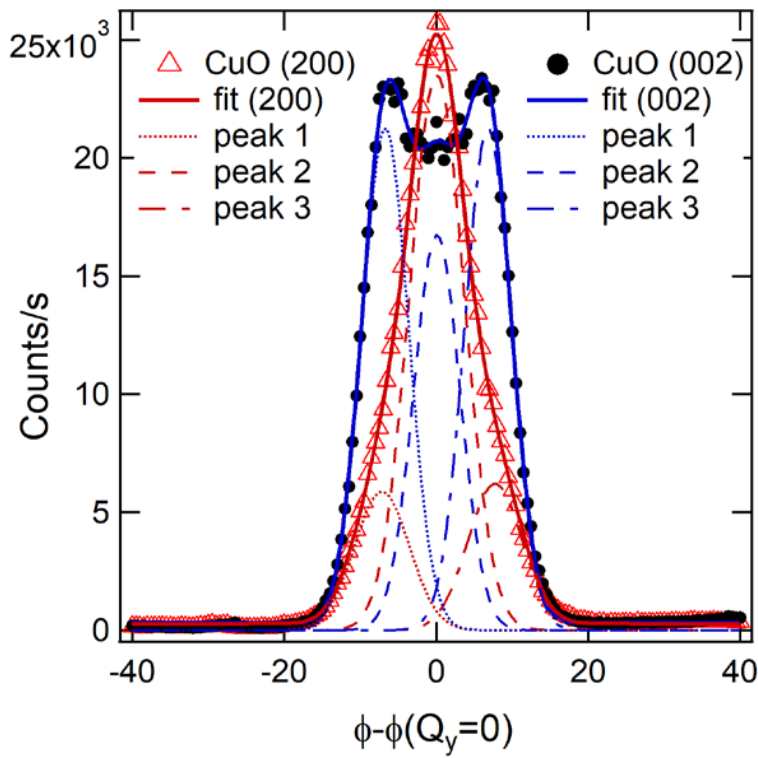

*Figure 7. Reciprocal space azimuthal scans performed at Q=24.84 nm$^{-1}$ and Q=27,26 nm$^{-1}$, corresponding respectively to the CuO(002) (full black circles) and CuO(200) reflections (empty*

*red triangles). They are centered on the STO [100] axis. Each reflection was fitted with three components, shown on the figure together with the best fit.*

The CuO(002) reflection (full black circles) shows two distinct peaks. However, the experimental intensity profile of this reflection can only be perfectly adjusted using three Gaussian components (see Table 2). In analogy, the Cu(200) reflection is also fitted with three components (Table 3). Combining these results, we see that indeed we have two kind of epitaxy, as schematized in Fig.8, giving by symmetry a total of 16 domains.

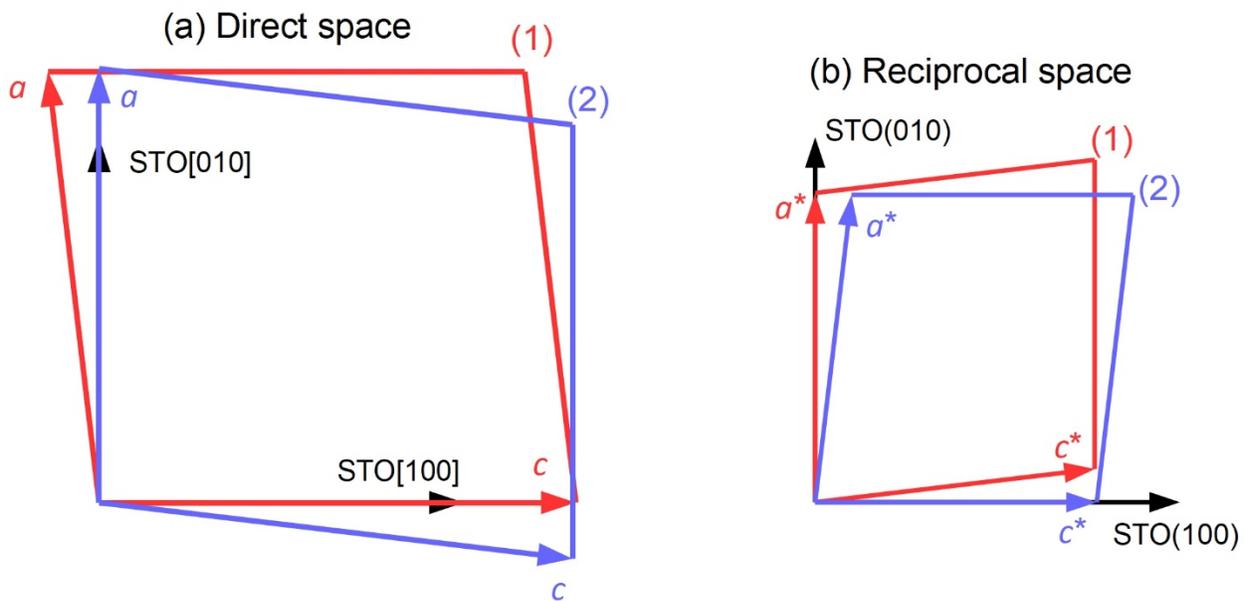

*Figure 8. (a) Direct and (b) reciprocal space schemas of the two CuO/STO epitaxy obtained from the XRD measurements.*

CuO[001]//STO[100], labeled (1) in figure 8, is the preferred orientation (more than 2/3 of intensity), but we also have domains with CuO[100]//STO[100], labeled (2). More important, the epitaxy relationship constrains the monoclinic angle $\beta$ to a value significantly smaller than the bulk one. On average, we obtain for epitaxy (1) $\beta_{film}=90°+(|\Phi_1|+|\Phi_2|)/2=96.76(8)°$ instead of $\beta_{bulk}=99.54°$.

**Table 2**. *Best fit of the Cu(002) azimuthal scan of fig. 7 with three Gaussian components. $\Phi$ represents the rotation with respect to the STO [100] direction.*

|   | $\Phi$ | I | FWHM |
|---|---|---|---|
| 1 | -6.79(4)° | 36% | 5.08(3)° |
| 2 | 6.73(4)° | 36% | - |
| 3 | 0.08(5)° | 28% | - |

**Table 3**. *Best fit of the Cu(200) azimuthal scan.*

|   | $\Phi$ | I | FWHM |
|---|---|---|---|
| 1 | 0.08(4)° | 66% | 6.01(6)° |
| 2 | -7.2(1)° | 16% | - |
| 3 | 7.7(1)° | 18% | - |

To crosscheck the film structure, we have finally scanned the momentum transfer perpendicular to the surface at $Q_x$ = 24.84 nm$^{-1}$ and $Q_x$ = 27.26 nm$^{-1}$ (corresponding respectively to the CuO(002) and CuO(200) peaks). Since all domains are oriented with the CuO [010] direction perpendicular to the surface, these scans have to cross the CuO(022) and CuO(220) peaks, respectively. These reflections are observed at the expected perpendicular momentum transfer, confirming the structural analysis (see Fig. 5 and Table 1).

## 4. Conclusions

In conclusion, we have investigated the structure at early growth stages of epitaxial CuO thin films by XRD using a synchrotron radiation source. The films were elaborated by post-deposition annealing. Cu was first deposited on SrTiO$_3$(001) in molecular oxygen at a partial pressure of 10$^{-4}$ Pa using molecular beam epitaxy. Afterwards, cupric oxide was obtained by annealing *ex situ* in molecular oxygen at atmospheric pressure. This elaboration method results, more commonly, in polycrystalline films, although there are examples of epitaxial films grown in this way. Thin films discussed above follow a Wolmer-Weber growth mode, i.e. are discontinuous and form epitaxial nanostructures a few tens of nanometer wide. The XRD analysis shows that the film structure is very close to the common CuO tenorite phase and that the [010] crystallographic axis is perpendicular to the STO surface. In bulk multiferroic phase the **b** axis is the direction of ferroelectric polarization. In the surface plane, islands are observed both with CuO[001]//STO[100] and CuO[100]//STO[100], the former being the preferred one. We also observe a significant distortion of the monoclinic $\beta$ angle induced by epitaxy, while the other lattice constants are bulk-like within the measurement's accuracy. Superexchange interaction in CuO is very sensitive to the Cu-O-Cu bonding angle, therefore each modification of the structural parameters is likely to modify its electronic and magnetic properties. Its ferroelectric properties and a high Néel temperature, together with a low structural symmetry, make tenorite an interesting candidate to improve the functionalities of technological devices. Knowledge of the processes that control the initial stages of growth of this system on crystalline substrates, such as SrTiO$_3$(001), is relevant for further understanding of the mechanisms that stabilize its properties and how these are modified as a function of the symmetry reduction imposed by the two-dimensionality of film growth.

**CRediT authorship contribution statement**

M. De Santis: Conceptualization, Investigation, Data analysis, Writing – original draft. V. Langlais and X. Torrelles: Investigation, Writing – review & editing. L. Martinelli and T. Mocellin: Investigation, S. Pairis: SEM Investigation.

**Data availability**

doi.esrf.fr/10.15151/ESRF-ES-778014448

**Acknowledgements**

We acknowledge the European Synchrotron Radiation Facility for provision of synchrotron radiation facilities and we would like to thank beamline staff for assistance in using beamline BM32 (French CRG IF). We thank Aude Bailly for helpful discussion and for a critical review of the manuscript. X.T., and V. L. acknowledge the public financial support of the CSIC ILINKA2021 Program through the ILINKA20394 project. X.T. also acknowledges grant PID2021-12327OB-I00 funded by MCIN/AEI/10.13039/501100011033 and by "ERDF A way of making Europe".